\documentclass[fleqn,usenatbib]{mnras}

\usepackage{mathptmx}

\usepackage[T1]{fontenc}

\DeclareRobustCommand{\VAN}[3]{#2}
\let\VANthebibliography\thebibliography
\def\thebibliography{\DeclareRobustCommand{\VAN}[3]{##3}\VANthebibliography}

\usepackage{graphicx}	
\usepackage{amsmath}	
\usepackage{amssymb}	

\title[Benchmarking MESA isochrones with the Hyades]{Benchmarking MESA isochrones against the Hyades single star sequence}

\author[W. Brandner, P. Calissendorff, and T. Kopytova]{
Wolfgang Brandner,$^{1}$\thanks{E-mail: brandner@mpia.de}
 Per Calissendorff$^{2}$
and Taisiya Kopytova$^{3,4,1}$
\\
$^{1}$Max-Planck-Institut f\"ur Astronomie, K\"onigstuhl 17, 69117 Heidelberg, Germany\\
$^{2}$Department of Astronomy, University of Michigan, Ann Arbor, MI 4810, USA\\
$^{3}$Division of Medical Image Computing, German Cancer Research Center (DKFZ), 69120 Heidelberg, Germany\\
$^{4}$Ural Federal University, Yekaterinburg, 620002, Russia
}

\date{Accepted 2022 August 5. Received 2022 July 29; in original form 2022 June 11}

\pubyear{2020}

\begin{document}
\label{firstpage}
\pagerange{\pageref{firstpage}--\pageref{lastpage}}
\maketitle

\begin{abstract}
Based on GAIA EDR3, we revisit and update our sample of bonafide single stars in the Hyades open cluster. The small observational uncertainties in parallax and photometry of EDR3 result in a tightly defined stellar sequence, which is ideal for the testing and calibration of theoretical stellar evolutionary tracks and isochrones. We benchmark the solar-scaled MESA evolutionary models against the single star sequence. We find that the non-rotating MESA models for [Fe/H] = +0.25 provide a good fit for stars with masses above 0.85, and very low mass stars below 0.25\,M$_\odot$. For stars with masses between 0.25 and 0.85\,M$_\odot$
the models systematically under predict the observed stellar luminosity. One potential limitation of the models for partially convective stars more massive than 0.35\,M$_\odot$ is the prescription of (superadiabatic) convection with the mixing-length theory parameter $\alpha_{\rm ML}$ tuned to match the Solar model. Below 0.35\,M$_\odot$, the increased scatter in the stellar sequence might be a manifestation of the {\it convective kissing instability}, which is driven by variations in the $^3$He nuclear energy production rate due to instabilities at the convective core to envelope boundary.
For a Hyades-like stellar population, the application of solar-scaled models to subsolar mass stars could result in a significant underestimate of the age, or an overestimate of the metallicity. We suggest that future grids of solar-scaled evolutionary stellar models could be complemented by Hyades-scaled models in the mass range 0.25 to 0.85\,M$_\odot$. 
\end{abstract}





\section{Introduction}

Stellar cluster and individual binary stars constitute the most important astrophysical calibration sources.
At an average distance of 45\,pc, the Hyades open cluster is the closest (populous) stellar cluster to the Sun. The Hyades have super-solar metallicity. \cite{Kopytova2016} derived [Fe/H]=+0.14 for the best-fitting BT-Settl2010+PISA  and DARTMOUTH isochrones \citep{Allard2013,Dotter2008,Tognelli2011,DaRio2012,Tognelli2012}, while \cite{Gossage2018} derived [Fe/H]=+0.10 to +0.12 using MESA isochrones \citep{Paxton2011,Paxton2013,Paxton2015,Dotter2016,Choi2016,Paxton2018}  in near infrared (NIR).

At an age of $\approx 635 \pm 135$\,Myr, the Hyades open cluster comprises main sequence and post-main sequence stars with initial masses in the range $\approx$0.1 to 3.6\,M$_\odot$ \citep[e.~g.][]{Perryman1998,deBruijne2001,Krumholz2019}, which serve as benchmarks for models of stellar evolution. Colour-absolute magnitude diagrams (CMD) \citep{Castellani2001,Roeser11} revealed discrepancies between stellar models and observations for sub-solar mass stars. \cite{Castellani2001} attributed this to limitations in the description of the efficiency of superadiabatic convection in the outer layers of partially convective stars. \cite{Kopytova2016} showed that the incorporation of updated input physics \citep[equation of state, opacities, etc.,][]{DeglInnocenti2008} results in an improved match between theoretical isochrones and 2MASS photometric measurements in the mass range 0.6 to 0.8\,M$_\odot$. Below 0.6\,M$_\odot$, the close-to-vertical (i.e.\ constant colour) stellar sequence in NIR CMDs makes isochronal fitting rather insensitive to stellar luminosity, and hence metallicity or age.

In \cite{Brandner2022} we used MESA and BHAC2015 \citep{Baraffe2015} isochrones in the {\it GAIA} photometric system to determine the age of the nearby exoplanet host star GJ~367. Both sets of isochrones suggested a young age in the range of $\approx$30 to 60\,Myr for the star, which is considerably younger than its age suggested by gyro-chronology, and by its space motion and galactic dynamics models. This and the unprecedented photometric and parallax accuracy of {\it GAIA} EDR3 observations prompted us to benchmark the solar-scaled MESA isochrones against the single star sequence of the Hyades open cluster. 

The structure of the paper is as follows. In section 2 we present the update sequence of bonafide single stars in the Hyades open cluster. In section 3 we summarize literature age and metallicity estimates of the Hyades based on photometric data. In section 4 we benchmark the MESA isochrones against the single star sequence in the {\it GAIA} photometric system. In section 5 we discuss potential short-comings of grids of stellar models, and suggest ways forward.


\section{The Hyades single star sequence}

\begin{table*}
\caption{Single and multiple candidate members of the Hyades cluster, classified according to {\it GAIA} EDR3, sorted by RA. The full table is available online.}             
\label{SingleStar}      
\centering                          
\begin{tabular}{r c c c c c c c c c c}        
GAIA EDR3 ID&  dpgeo & lo$\_$dpgeo & hi$\_$dpgeo & G & $\sigma_{\rm G}$& BP & $\sigma_{\rm BP}$ & RP & $\sigma_{\rm RP}$ & flag \\
            & [pc]     & [pc]      & [pc]       & [mag]&[mag]& [mag]&[mag]&[mag]&[mag]& \\
\hline                                   
    395696646953688448& 59.612 & 59.565 & 59.655 &10.7791& 0.0009& 11.4353& 0.0004& 10.0031& 0.0002&      1\\
    386277165192421120& 30.935 & 30.908 & 30.962 &14.7670& 0.0089& 16.6299& 0.0030& 13.4805& 0.0009&      2\\
    393017579491591168& 64.815 & 64.423 & 65.115 &17.0372& 0.0073& 19.2764& 0.0380& 15.6597& 0.0031&      1\\
    420637590762193792& 83.610 & 82.524 & 84.646 &18.5022& 0.0007& 20.9101& 0.0994& 17.0266& 0.0068&      1\\
    385502112574538624& 42.821 & 42.780 & 42.859 &14.3729& 0.0005& 16.1299& 0.0041& 13.1075& 0.0015&      1\\
   2860677398591440768& 39.751 & 39.395 & 40.054 &11.7972& 0.0017& 12.8881& 0.0017& 10.6002& 0.0022&      4\\
\end{tabular}
    \begin{quote}
        flag: 1 - bonafide single, 2 - likely binary or multiple, 3 - white dwarf, 4 - peculiar {\it GAIA} EDR3 BP-G vs.\ G-RP colours;  Median , low (lo, 16th quantile) and high (hi, 84th quantile) of the photogeometric distance posterior dpgeo are from \cite{Bailer2021}
      \end{quote}
\end{table*}

\begin{figure*}
    \begin{center}
        \includegraphics[width=1.0\textwidth]{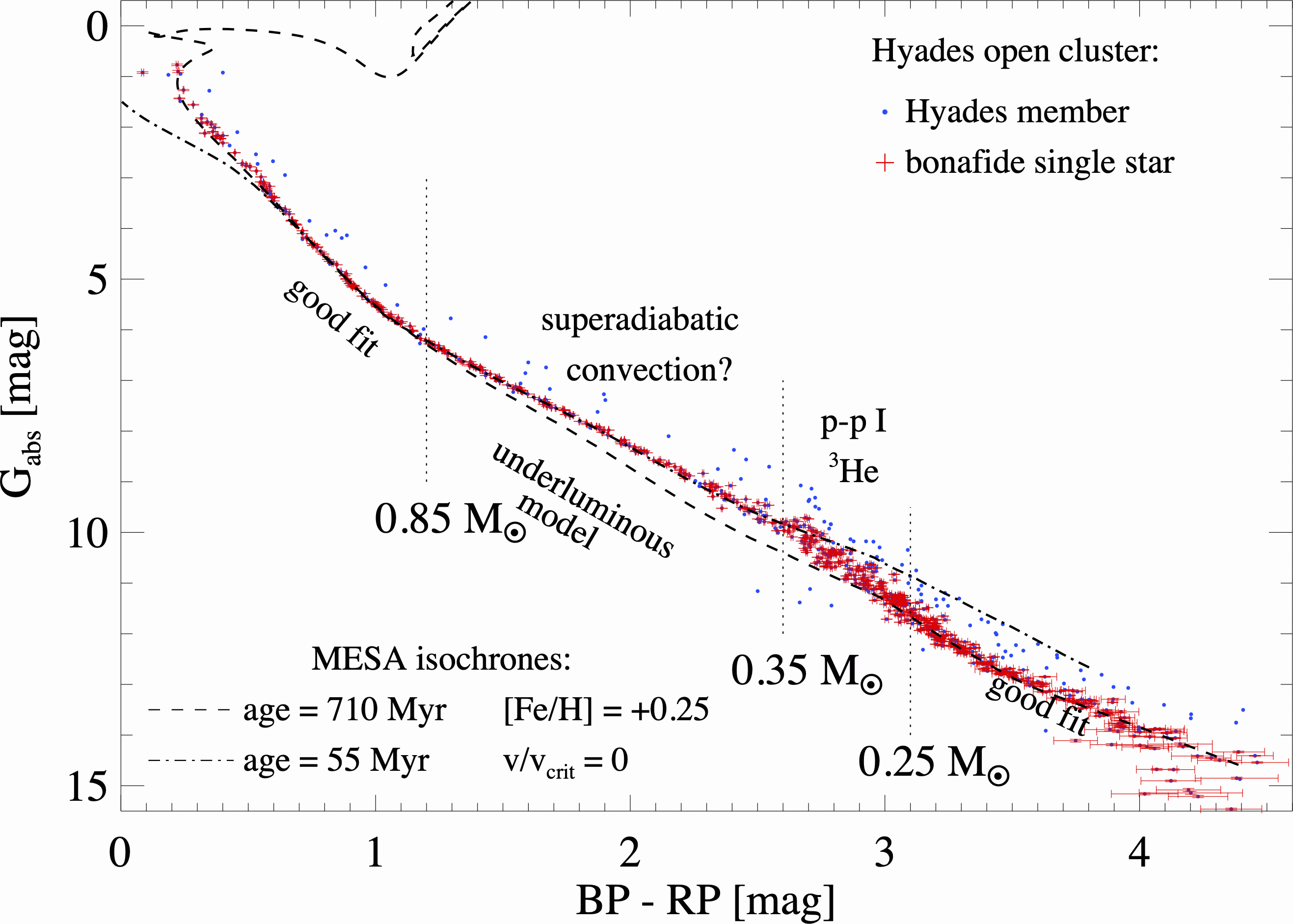}
    \end{center}
    \caption[]{Colour-absolute magnitude diagram of members of the Hyades open cluster based on GAIA EDR3 parallaxes and photometry. Blue dots indicate all stars in the GAIA GCNS sample with good photometry. Red crosses indicate members of our bonafide single star sequence. Overplotted are MESA isochrones. While the isochrone matching the age of the Hyades ($\approx$710\,Myr, dashed line) provides a good fit to the upper and lower main sequence, it underpredicts the luminosity of stars in the mass range $\approx$0.25 to 0.85\,M$_\odot$. The 1$\sigma$ error bars for the single stars are based on the uncertainties in photometry and parallax.}
    \label{CMDfull}
\end{figure*}

\cite{Kopytova2016} defined a fiducial observational sequence of single stars suitable for testing of stellar evolutionary and atmospheric models. The stars were selected from a sample of 724 probable members of the Hyades open cluster established by \cite{Roeser11} based on their proper motion according to the PPMXL catalog \citep{Roeser2010}. Using literature data and high-angular resolution Lucky Imaging observations with AstraLux Norte \citep{Hormuth2008}, the stars were screened for stellar binarity and photometric blends to derive a sample of single stars. This single star sample was selected quite conservatively. Considering the intrinsic 2$''$ angular resolution of the 2MASS Point Source Catalog \citep{Cutri2003}, and in order to minimize the effect of photometric blends, \cite{Kopytova2016} flagged all occurrences of another source within $\approx$4$''$ as potential binary companions, and excluded them from the single star sample. 

{\it GAIA} EDR3 facilitates a refinement of the single star sequence from \cite{Kopytova2016}. \cite{GAIA_Smart2021A} published a {\it GAIA} Catalogue of Nearby Stars (GCNS) listing 920 candidate members of the Hyades. In order to reject photometric outliers caused, e.g., by blends in the {\it GAIA} BP and RP bands, we first applied a colour cut-off: $-0.2$\,mag $\le$ BP - G $\le$ 3.2\,mag and  $-0.3$\,mag $\le$ G - RP $\le$ 1.7\,mag. This rejected 30 sources, resulting in a sample of 890 candidate members listed in Table \ref{SingleStar}. As a second step we fitted a 4th order polynomial to the data in a G - RP vs.\ BP - G two-colour diagram.  Application of an iterative sigma-clipping resulted in a sample of 783 candidate members with good photometric quality data. As a third step, we used the Renormalized Unit Weight Error (RUWE, see \cite{gaia_edr3lite,Lindegren2021}) to distinguish between bonafide single stars and likely unresolved binary and multiple systems. RUWE values around 1.0 indicate that the {\it GAIA} astrometric observations are well fitted by the single-star model. A significantly larger RUWE value indicates that the single-star model does not provide a good fit to the astrometric solution due to, e.g., the non-single nature of the source.

Next we computed absolute G$_{\rm abs}$ magnitudes based on the apparent G magnitudes and the EDR3 photogeometric distances according to \cite{Bailer2021}\footnote{For 18 stars, including five of the Hyades white dwarfs (GAIA EDR3 ID 45980377978968064, 3313606340183243136, 3313714023603261568, 3306722607119077120, 3294248609046258048), we substituted the missing photogeometric distance by the geometric distance.}. In order to identify and flag photometric binaries (i.e.\ sources falling on the binary sequence in the CMD, but with separations too close to be identified by the RUWE selection), we fitted an 8th order polynomial to the single star main-sequence, and applied an iterative sigma clipping. This resulted in 616 sources classified as bonafide single stars, 156 as likely binary or multiple systems, and 11 as white dwarfs.

\begin{figure*}
    \begin{center}
        \includegraphics[width=1.0\textwidth]{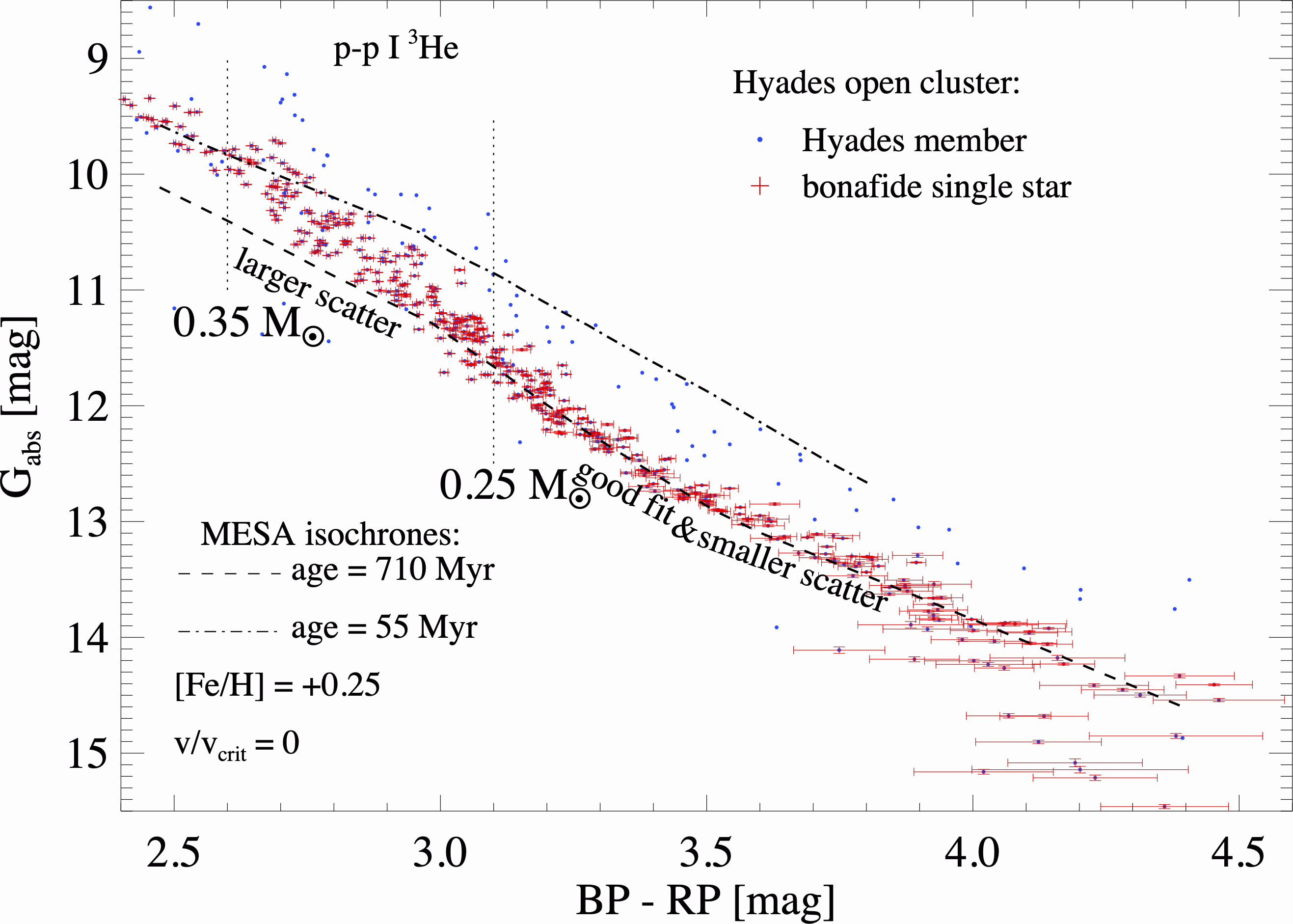}
    \end{center}
    \caption[]{Same as Figure \ref{CMDfull}, but zoomed in on very low-mass stars with masses $\lessapprox$0.35\,M$_\odot$.}
    \label{CMDvlm}
\end{figure*}

Figure \ref{CMDfull} shows the colour-absolute magnitude diagram of the Hyades open cluster, covering the main sequence and part of the post-main sequence. Blue dots mark the candidate members of the Hyades from the GCNS sample. Red crosses mark bonafide single stars, with observational uncertainties derived from uncertainties in {\it GAIA} photometry and parallax indicated. The median uncertainty in BP-RP colour amounts to 4.1\,mmag, and to 3.3\,mmag in G$_{\rm abs}$.
In particular for BP-RP $\le 2.3$\,mag, the single stars form a very tight sequence, which is clearly distinct from the scatter of apparently overluminous Hyades members located on the binary sequence. For redder (and intrinsically fainter) stars, there is a larger scatter in the bonafide single star sequence.

As Figure \ref{CMDvlm} highlights, the single star sequence becomes successively incomplete for BP-RP $\ge 3.2$\,mag. The sample of GCNS Hyades candidate members extends to BP-RP = 4.7\,mag. The reddest and lowest mass object included in the single star sample is LSPM J0354+2316 ({\it GAIA} EDR3 65638443294980224), which is of spectral type M8 \citep{Bardalez2014}, and has a mass of $\approx$0.1\,M$_\odot$ \citep{Goldmann2013}. 


\section{Age and metallicity of the Hyades}

There is a vast literature on abundance estimates and the calibration of astrophysical parameters for stars in the Hyades cluster (see, e.g., \cite{Perryman1998,Tognelli2021}, and references therein). Age estimates for the Hyades are in general derived from the main sequence turn-off and isochrone fitting. Abundance estimates rely both on spectral analysis and isochrone fitting. 

\begin{table*}
\caption{Compilation of abundance, $\alpha_{\rm ML}$, and age estimates for the Hyades based on isochrone fitting}             
\label{hyades_metal}      
\centering                          
\begin{tabular}{l c c c c c c c c l}        
[Fe/H] & X & Y & Z& $\Delta$Y/$\Delta$Z& $\alpha_{\rm ML}$ & mass range & age  &PD$^1$&reference \\
 &   &   &  & &  & [M$_\odot$] & [Myr] & & \\ \hline
$+0.14${\raisebox{0.5ex}{\tiny$^{+0.05}_{-0.05}$}} &0.716 & $0.260${\raisebox{0.5ex}{\tiny$^{+0.020}_{-0.020}$}} & $0.024${\raisebox{0.5ex}{\tiny$^{+0.003}_{-0.003}$}}& &1.64 &[0.8,1.6] & $625\pm 50$ &BD&\cite{Perryman1998}\\
+0.14         &0.691     &0.285  &0.024    & &1.68  &[0.5,0.9]  &$638 \pm 13$ &TY&\cite{deBruijne2001}\\
+0.14         &0.716     &0.260  &0.024    & &1.64  &[0.9,1.6] &$638 \pm 13$ &TY&\cite{deBruijne2001}\\
+0.14         &0.708     &0.273  &0.019    & &1.68  &[1.6,2.4]  &$631$ &TY&\cite{deBruijne2001}\\
$+0.14$       &0.700  &0.283  &0.0175  &2 &1.74  &[0.13,2.30]  &$726 \pm 50$ &2M&\cite{Kopytova2016}\\
$+0.24${\raisebox{0.5ex}{\tiny$^{+0.02}_{-0.02}$}}&  &  &  & &1.82 &[0.5,2.4]  &$726 \pm 50$ &TY&\cite{Gossage2018}\\
$+0.10${\raisebox{0.5ex}{\tiny$^{+0.02}_{-0.02}$}}&  &  &  & &1.82&[0.5,2.4]  &741{\raisebox{0.5ex}{\tiny$^{+36}_{-14}$}} &2M&\cite{Gossage2018}\\
$+0.169${\raisebox{0.5ex}{\tiny$^{+0.025}_{-0.025}$}} &0.6947 &0.2867 &0.01863 &$2.03\pm0.33$ &$2.01\pm0.05$ &[0.83,1.35]&500&G2& \cite{Tognelli2021}\\ \hline
\end{tabular}
    \begin{quote}
        $^1$ key to photometric data set (PD): 2M - based on 2MASS photometry \citep{Cutri2003}; BD - based on BDA \citep{Mermilliod1995}; G2 - based on {\it GAIA} DR2 \citep{GAIA2016,GAIA2018}; TY - based on {\it TYCHO} photometry \citep{Hog2000} 
      \end{quote}
\end{table*}

Table \ref{hyades_metal} summarizes some of the canonical estimates, including the helium-to-metal enrichment ratio $\Delta$Y/$\Delta$Z, based on isochrone fitting. The differences in the parameter estimates can in part be explained by variations in the observational methods and data sets and their intrinsic uncertainties, in part by differences and advances in the modelling of stellar interiors and atmospheres, and in part by advances in the analysis of the solar elemental abundances (see, e.g., \cite{Asplund2009}). The majority of the estimates focused on post-main sequence stars and main sequence stars of spectral type K and earlier (more massive than 0.5\,M$_\odot$). The sole exception is the study by \cite{Kopytova2016}, which includes stars with masses as low as 0.13\,M$_\odot$. Some of the studies also consider variations of the mixing length (ML) parameter $\alpha_{\rm ML}$ or stellar rotation \citep{Gossage2018,Tognelli2021}. The latter effect appears to be most noticeable in the colours and luminosity of post-main sequence stars.

Common to all studies is the derived (or assumed) super-solar metallicity of the Hyades, with [Fe/H] estimates in the range $+0.10$ to $+0.24$. Age estimates for the Hyades cover the range 500 to 770\,Myr. The majority of the studies also indicate a higher than solar He abundance for the Hyades (see Table \ref{hyades_metal}).


\section{Benchmarking isochrones}

\begin{figure*}
    \begin{center}
        \includegraphics[width=1.0\textwidth]{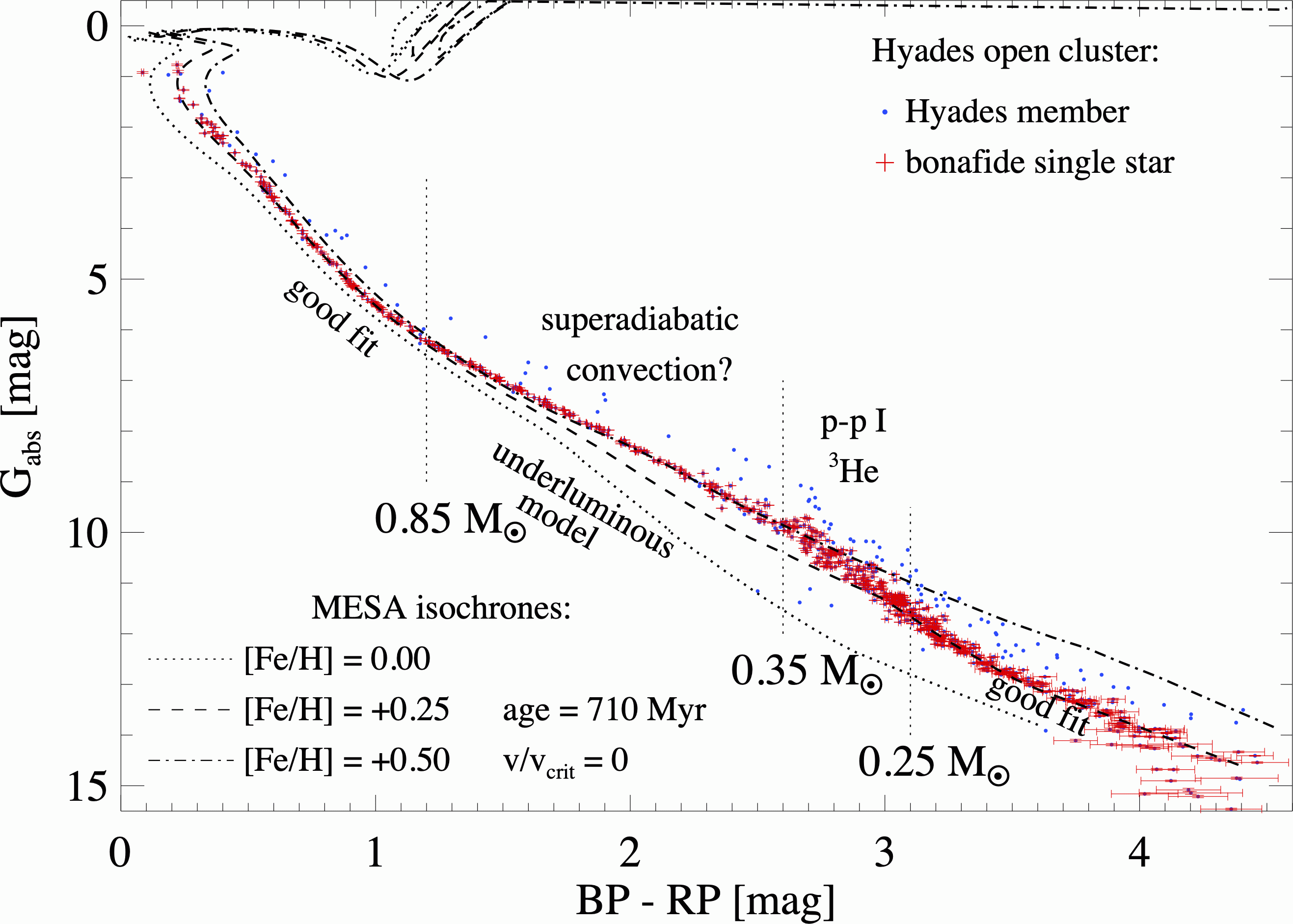}
    \end{center}
    \caption[]{Same as Figure \ref{CMDfull}, but varying the stellar metallicity of the 710\,Myr isochrones.}
    \label{CMDfull_metal}
\end{figure*}

\begin{table*}
\caption{Model stellar parameters at inflection points between MESA isochrone (age = 710 Myr, [Fe/H]=+0.25, v/v$_{\rm crit}$ = 0) and Hyades single star sequence}             
\label{StellarColourRegions}      
\centering                          
\begin{tabular}{c c c c c c c c}        
G$_{\rm abs}$$^1$& BP - RP$^1$&  B - V & log T$_{\rm eff}$ & log g & log L &  Mass & note \\

 [mag]& [mag]& [mag]&  [K]&  [cm/s$^2$]& [L$_\odot$] &  [M$_\odot$] &  \\ \hline
$<$6.7&$<$1.2      &$<$1.03      &$>$3.676       &$<$4.61 &$>$-0.594          &$>$0.85  &good fit\\
6.7 to 11.2 &1.2 to 2.6  &1.03 to 1.30 &3.676 to 3.519 &4.61 to 4.90 &-0.594 to -1.864 &0.85 to 0.35&model underluminous\\
11.2 to 13.5&2.6 to 3.1  &1.30 to 1.36 &3.519 to 3.485 &4.90 to 4.97 &-1.864 to -2.229 &0.35 to 0.25& transition region \\
$>$13.5&$>$3.1      &$>$1.36      &$<$3.485       &$>$4.97 &$<$-2.229        &$\le$0.25 &good fit\\
\hline                                   
\end{tabular}
    \begin{quote}
        $^1$ G$_{\rm abs}$ and BP-RP refer to the observed stellar sequence, while the other quantities are according to the MESA isochrone for the corresponding BP-RP colour.
      \end{quote}
\end{table*}

The solar scaled MESA isochrones and stellar tracks\footnote{We use the \detokenize{MIST_v1.2_vvcrit0.0_UBVRIplus} packaged model grid, dated 2020-12-04, which includes updated synthetic photometry for {\it GAIA} EDR3 based on \cite{Riello2021}. The grid steps are 0.25 dex in the range -2.00 $\le $ [Fe/H] $\le$+0.50, and 0.05 dex in the range 5.0 $\le \log_{10}$(age [yr]) $\le$ 10.3} use a Ledoux plus mixing length theory prescription of convection, with $\alpha_{\rm ML} = 1.82$ tuned to fit the Sun \citep{Choi2016}.

In Figure \ref{CMDfull} we overlay two MESA isochrones for [Fe/H]=+0.25, and no rotation (v/v$_{\rm crit}$ = 0) on the colour-absolute magnitude diagram of the Hyades. As presented in \cite{Gossage2018}, the best fitting MESA isochrones in the optical yield systematically higher metallicities than the best fitting MESA isochrones in the NIR for the Hyades, Praesepe, and Pleiades clusters.\footnote{\cite{Gossage2018} attribute this to the small number of stars in their optical samples ($<$40 stars for the Hyades, according to their Figure 7), and suggest that their optical samples do not provide meaningful constraints on the metallicity of either of these clusters.}
In the pre-computed grid of solar-scaled MESA isochrones, [Fe/H]=+0.25 is closest to [Fe/H]=+0.24$\pm$0.01 as deduced by \cite{Gossage2018} from the best fitting MESA isochrone in the {\it TYCHO} B$_{\rm T}$ ,V$_{\rm T}$ photometric system. The choice of non-rotating stellar models is based on the dearth of rapid rotators among single stars with masses $\ge$0.3\,M$_\odot$ in the Hyades \citep{Douglas2016}. They find that stars with masses of $\approx$0.4\,M$_\odot$ from the sample defined by \cite{Kopytova2016} have typical rotational periods of $\approx$20\,days.

We find that the isochrone for $\log_{10}$(age [yr]) = 8.85 ($\approx$710\,Myr) provides a better fit for stars with masses $>$1.35\,M$_\odot$ than the next younger (630\,Myr) or older isochrones (795\,Myr). This age is in good agreement with isochronal age determinations by \cite{Kopytova2016} and \cite{Gossage2018}. \cite{Tognelli2021} only considered stars with masses $<$1.5\,M$_\odot$ for the age determination, and were thus less sensitive to the rapid evolution of stellar luminosity near the upper end of the main sequence.
For stars between 0.25 and 0.85\,M$_\odot$, the 710\,Myr isochrone tends to underpredict the stellar luminosity. In the colour range BP-RP = 1.2 to 2.6\,mag, the 55\,Myr isochrone (dash-dotted line) provides a good fit to the observed sequence, but it overpredicts the stellar luminosity for BP-RP $\ge$2.6\,mag.

In Figure \ref{CMDfull_metal} we overlay three isochrones for an age of 710\,Myr, and for [Fe/H] = 0.00, +0.25, and +0.50. The highest metallicity isochrone (dash-dotted line) provides a good fit the colour range BP-RP = 1.2 to 2.6\,mag, but overpredicts the stellar luminosity for bluer and redder stars.

Table \ref{StellarColourRegions} lists the corresponding B-V colour, effective temperature, $\log$ g, $\log$ L, and stellar mass according to the MESA isochrone (age = 710\,Myr ([Fe/H]=+0.25, and v/v$_{\rm crit}$ = 0) at the boundaries of the four BP-RP colour regions marked by the vertical dotted lines in Figures \ref{CMDfull} and \ref{CMDfull_metal}. None of the single-age, single-metallicity isochrones is capable of fitting the entire single star sequence.


\section{Discussion}

In Figures \ref{CMDfull} and \ref{CMDfull_metal} we have marked the BP-RP colour regions where the 710\,Myr, [Fe/H] = +0.25 isochrone provides a good fit to the observed sequence, and where it significantly deviates from the observed sequence by predicting fainter (underluminous) stars. 
The good fit of the MESA isochrones for solar-type stars with masses above 0.85\,M$_\odot$, and for very low-mass stars with masses between 0.1 and 0.25\,M$_\odot$ is highly encouraging, and speaks for the maturity of stellar model grids. 

Potential problem areas in modelling CMDs of the Hyades have been noticed in the literature. In general, the challenges considered for the Hyades stellar sequence are related to stellar rotation, elemental abundance ([Fe/H] and $\Delta {\rm Y}/\Delta {\rm Z}$) and nuclear energy production rates, and the description of convection. \cite{Gossage2018} discusses the effect of rotation and variations in $\alpha_{\rm ML}$ for stars above 1.2\,M$_\odot$ in the Hyades. They conclude that a larger $\alpha_{\rm ML} = 2.0$ would provide a better fit to the giant stars in the Hyades. For Hyades members with masses less than 0.85\,M$_\odot$, \cite{Castellani2001} noted a discrepancy between observed and theoretical optical colour and brightness, in particular in the mass range where superadiabatic convection dominates the outer convective zone of a star. They suggest to consider $\alpha_{\rm ML}$ as a free parameter in this mass range which could be tuned to a (lower) value to better describe the efficiency of superadiabatic convection. Based in part on the ideas initially explored by \cite{Stevenson1979} and \cite{MacDonald2014}, \cite{Ireland2018} study the effects of rotation and magnetic fields on convection, and investigate how they could be parameterized by a depth dependent $\alpha_{\rm ML}$.

The good match of the synthetic photometry of the MESA isochrone to the observed GAIA data for BP-RP $>$3.1\,mag suggests that there is no generic issue in the ATLAS12 and SYNTHE conversion \citep{Choi2016} of luminosity and temperature to synthetic photometry for cool stellar photospheres. \cite{Choi2016} use different sets of opacity tables for the mass ranges 0.1 to 0.3\,M$_\odot$, 0.3 to 0.6\,M$_\odot$, and $>$0.6\,M$_\odot$ as boundary conditions  for the atmospheres. The discrepancy for 1.2\,mag$<$BP-RP$<$3.1\,mag (T$_{\rm eff}$ = 3050 to 4750\,K, m = 0.25 to 0.85\,M$_\odot$), thus might warrant a review of the opacity tables and their transitions in this parameter range. \cite{Choi2016} also point out systematic differences in the evolutionary track of a 0.3\,M$_\odot$ star between the MIST and Lyon \citep{Baraffe1998,Baraffe2003,Baraffe2015} models on the one side, and the PARSEC \citep{Giradi2002,Marigo2008,Bressan2012} models on the other side. They attribute this difference to the modified temperature-Rosseland mean optical depth (T-$\tau$) relation \citep{Chen2014} employed for low-mass stars by the PARSEC models.

The colour range around BP-RP $\approx$2.80 to 2.95\,mag (G$_{\rm abs} \approx 10.5$ to 11\,mag) stands out in the CMD as the observed stellar sequence shows a larger scatter than for stars with bluer colours or BP-RP $>$ 3.1\,mag (G$_{\rm abs} > 11.0$\,mag, Figures \ref{CMDfull} and \ref{CMDvlm}). According to the MESA tracks, this correspond to stellar masses just below 0.35\,M$_\odot$, which roughly coincides with the fully convective boundary. As discussed by \cite{Baraffe2018}, for stars in this mass range the energy productions rate of the proton-proton I branch ($^3$He + $^3$He $\longrightarrow$ $^4$He + 2 p) is crucial for a proper description of the stellar luminosity. Depending on the precise stellar properties and the initial He abundance, $^3$He in Hyades members in this mass range might not yet have reached its equilibrium abundance. An overabundance in $^3$He resulting in an enhanced energy productions rate could explain the observed overluminosity of stars compared to the 710\,Myr isochrone. The increased scatter in the stellar magnitude-colour sequence could suggest the presence of an instability in the stellar luminosity for this particular mass and age range. \cite{vanSaders2012} identified a $^3$He-driven instability for stars near the fully convective boundary, which they referred to as {\it convective kissing instability}. \cite{Baraffe2018} confirmed the existence of this instability using a different evolutionary code. Caveats are that the observed increase in the scatter of G$_{\rm abs}$ by $\approx$0.2\,mag (20\%) is about a factor 2 to 4 larger than the variations in luminosity according to the models by \cite{vanSaders2012}. For stellar interior models of solar metallicity, the convective kissing instability seems to be restricted to a relatively narrow mass range of 0.34 to 0.37\,M$_\odot$ \citep{vanSaders2012,Baraffe2018}.

In the Hyades, a stellar mass of $\approx$0.30\,M$_\odot$ marks the boundary between very low mass stars with rotational periods ranging from a fraction of a day to 5 days, and more massive single stars with rotational periods in the range of 10 to 20\,days \citep{Douglas2016,Douglas2019}. Fast rotation and the associated stellar activity has been associated with radius inflation \citep{Somers2015a,Somers2017}. As discussed by \cite{Feiden2014} and \cite{Feiden2016} a significant radius inflation requires strong interior magnetic fields in the range of 10\,MG, which might be difficult to maintain over the age of the Hyades. \cite{Douglas2016} discuss that poloidal fields in Hyades members in the mass range 0.3 to 0.6\,M$_\odot$ resulted in effective magnetic braking, and strongly reduced stellar activity, while fully convective stars of lower mass primarily rely on their (weak) stellar winds for shedding angular momentum. We suggest that a future study could focus on the rotation periods and activity levels of the stars in the mass range 0.25 and 0.35\,M$_\odot$, and look for, e.g., correlations with their luminosity, or photometric variability.

As of mid 2022, \cite{Choi2016} had more than 1200 citations, with the MESA models being used to assess ages, metallicity, radii, effective temperatures, luminosity, surface gravity, etc.\ for individual stars and (complex) stellar populations. 
The primary science of more than 130 of these articles is on exoplanets, where in general planetary properties are derived relative to the astrophysical properties of the host star. Biased stellar properties could thus directly bias the deduced properties of exoplanets.

The example of the Hyades single star sequence highlights the potential perils in the analysis of low-mass (late-type) stellar populations. Even in the presence of {\it GAIA} high-precision parallax and photometric information, missing supplemental information could result in biased conclusions on absolute stellar ages, metallicity, or the intrinsic spread of these properties. A `blind' isochronal analysis of the Hyades sample in the BP-RP colour range 1.2 to 2.6\,mag might underestimate the true age by more than a factor of 10 (55 vs 710\,Myr, see Figure \ref{CMDfull}), or result in a significant overestimate of its metallicity ([Fe/H] = +0.50 vs. +0.25, see Figure \ref{CMDfull_metal}).

The updated single star sequence of the Hyades cluster with its accurate {\it GAIA} EDR3 distance and photometric measurements could serve as a reference to tune modelling parameters like, e.g., $\alpha_{\rm ML}$ to the efficiency of super-adiabatic convection, or to tune astrophysical parameters like, e.g., $\Delta {\rm Y}/\Delta {\rm Z}$ (and in particular $^3$He abundances) to reflect actual energy productions rates. 
We suggest that future grids of solar-scaled evolutionary models should be tested against the Hyades single star sequence presented in Table \ref{SingleStar}, or against comparable data sets for the Pleiades or Praesepe open clusters.


\section*{Acknowledgements}

We thank H.-W.\ Rix for the initial discussion, which prompted this research. We thank the anonymous referee for constructive comments, which helped to improve the paper.

This work has made use of data from the European Space Agency (ESA) mission
{\it Gaia} (\url{https://www.cosmos.esa.int/gaia}), processed by the {\it Gaia}
Data Processing and Analysis Consortium (DPAC,
\url{https://www.cosmos.esa.int/web/gaia/dpac/consortium}). Funding for the DPAC
has been provided by national institutions, in particular the institutions
participating in the {\it Gaia} Multilateral Agreement.

\section*{Data availability}
The data underlying this article are available in the article and in its online supplementary material. The online version of Table \ref{SingleStar} includes coordinates, which makes objects discoverable in VizieR.



\bibliographystyle{mnras}
\bibliography{lit} 

\begin{thebibliography}{}
\makeatletter
\relax
\def\mn@urlcharsother{\let\do\@makeother \do\$\do\&\do\#\do\^\do\_\do\%\do\~}
\def\mn@doi{\begingroup\mn@urlcharsother \@ifnextchar [ {\mn@doi@}
  {\mn@doi@[]}}
\def\mn@doi@[#1]#2{\def\@tempa{#1}\ifx\@tempa\@empty \href
  {http://dx.doi.org/#2} {doi:#2}\else \href {http://dx.doi.org/#2} {#1}\fi
  \endgroup}
\def\mn@eprint#1#2{\mn@eprint@#1:#2::\@nil}
\def\mn@eprint@arXiv#1{\href {http://arxiv.org/abs/#1} {{\tt arXiv:#1}}}
\def\mn@eprint@dblp#1{\href {http://dblp.uni-trier.de/rec/bibtex/#1.xml}
  {dblp:#1}}
\def\mn@eprint@#1:#2:#3:#4\@nil{\def\@tempa {#1}\def\@tempb {#2}\def\@tempc
  {#3}\ifx \@tempc \@empty \let \@tempc \@tempb \let \@tempb \@tempa \fi \ifx
  \@tempb \@empty \def\@tempb {arXiv}\fi \@ifundefined
  {mn@eprint@\@tempb}{\@tempb:\@tempc}{\expandafter \expandafter \csname
  mn@eprint@\@tempb\endcsname \expandafter{\@tempc}}}

\bibitem[\protect\citeauthoryear{{Allard}, {Homeier}  \& {Freytag}}{{Allard}
  et~al.}{2013}]{Allard2013}
{Allard} F.,  {Homeier} D.,   {Freytag} B.,  2013, \memsai, \href
  {https://ui.adsabs.harvard.edu/abs/2013MmSAI..84.1053A} {84, 1053}

\bibitem[\protect\citeauthoryear{{Asplund}, {Grevesse}, {Sauval}  \&
  {Scott}}{{Asplund} et~al.}{2009}]{Asplund2009}
{Asplund} M.,  {Grevesse} N.,  {Sauval} A.~J.,   {Scott} P.,  2009, \mn@doi
  [\araa] {10.1146/annurev.astro.46.060407.145222}, \href
  {https://ui.adsabs.harvard.edu/abs/2009ARA&A..47..481A} {47, 481}

\bibitem[\protect\citeauthoryear{{Bailer-Jones}, {Rybizki}, {Fouesneau},
  {Demleitner}  \& {Andrae}}{{Bailer-Jones} et~al.}{2021}]{Bailer2021}
{Bailer-Jones} C.~A.~L.,  {Rybizki} J.,  {Fouesneau} M.,  {Demleitner} M.,
  {Andrae} R.,  2021, \mn@doi [\aj] {10.3847/1538-3881/abd806}, \href
  {https://ui.adsabs.harvard.edu/abs/2021AJ....161..147B} {161, 147}

\bibitem[\protect\citeauthoryear{{Baraffe} \& {Chabrier}}{{Baraffe} \&
  {Chabrier}}{2018}]{Baraffe2018}
{Baraffe} I.,  {Chabrier} G.,  2018, \mn@doi [\aap]
  {10.1051/0004-6361/201834062}, \href
  {https://ui.adsabs.harvard.edu/abs/2018A&A...619A.177B} {619, A177}

\bibitem[\protect\citeauthoryear{{Baraffe}, {Chabrier}, {Allard}  \&
  {Hauschildt}}{{Baraffe} et~al.}{1998}]{Baraffe1998}
{Baraffe} I.,  {Chabrier} G.,  {Allard} F.,   {Hauschildt} P.~H.,  1998, \aap,
  \href {https://ui.adsabs.harvard.edu/abs/1998A&A...337..403B} {337, 403}

\bibitem[\protect\citeauthoryear{{Baraffe}, {Chabrier}, {Barman}, {Allard}  \&
  {Hauschildt}}{{Baraffe} et~al.}{2003}]{Baraffe2003}
{Baraffe} I.,  {Chabrier} G.,  {Barman} T.~S.,  {Allard} F.,   {Hauschildt}
  P.~H.,  2003, \mn@doi [\aap] {10.1051/0004-6361:20030252}, \href
  {https://ui.adsabs.harvard.edu/abs/2003A&A...402..701B} {402, 701}

\bibitem[\protect\citeauthoryear{{Baraffe}, {Homeier}, {Allard}  \&
  {Chabrier}}{{Baraffe} et~al.}{2015}]{Baraffe2015}
{Baraffe} I.,  {Homeier} D.,  {Allard} F.,   {Chabrier} G.,  2015, \mn@doi
  [\aap] {10.1051/0004-6361/201425481}, \href
  {https://ui.adsabs.harvard.edu/abs/2015A&A...577A..42B} {577, A42}

\bibitem[\protect\citeauthoryear{{Bardalez Gagliuffi} et~al.,}{{Bardalez
  Gagliuffi} et~al.}{2014}]{Bardalez2014}
{Bardalez Gagliuffi} D.~C.,  et~al., 2014, \mn@doi [\apj]
  {10.1088/0004-637X/794/2/143}, \href
  {https://ui.adsabs.harvard.edu/abs/2014ApJ...794..143B} {794, 143}

\bibitem[\protect\citeauthoryear{{Brandner}, {Calissendorff}, {Frankel}  \&
  {Cantalloube}}{{Brandner} et~al.}{2022}]{Brandner2022}
{Brandner} W.,  {Calissendorff} P.,  {Frankel} N.,   {Cantalloube} F.,  2022,
  \mn@doi [\mnras] {10.1093/mnras/stac961}, \href
  {https://ui.adsabs.harvard.edu/abs/2022MNRAS.tmp..926B} {513, 661}

\bibitem[\protect\citeauthoryear{{Bressan}, {Marigo}, {Girardi}, {Salasnich},
  {Dal Cero}, {Rubele}  \& {Nanni}}{{Bressan} et~al.}{2012}]{Bressan2012}
{Bressan} A.,  {Marigo} P.,  {Girardi} L.,  {Salasnich} B.,  {Dal Cero} C.,
  {Rubele} S.,   {Nanni} A.,  2012, \mn@doi [\mnras]
  {10.1111/j.1365-2966.2012.21948.x}, \href
  {https://ui.adsabs.harvard.edu/abs/2012MNRAS.427..127B} {427, 127}

\bibitem[\protect\citeauthoryear{{Castellani}, {Degl'Innocenti}  \& {Prada
  Moroni}}{{Castellani} et~al.}{2001}]{Castellani2001}
{Castellani} V.,  {Degl'Innocenti} S.,   {Prada Moroni} P.~G.,  2001, \mn@doi
  [\mnras] {10.1046/j.1365-8711.2001.03958.x}, \href
  {https://ui.adsabs.harvard.edu/abs/2001MNRAS.320...66C} {320, 66}

\bibitem[\protect\citeauthoryear{{Chen}, {Girardi}, {Bressan}, {Marigo},
  {Barbieri}  \& {Kong}}{{Chen} et~al.}{2014}]{Chen2014}
{Chen} Y.,  {Girardi} L.,  {Bressan} A.,  {Marigo} P.,  {Barbieri} M.,   {Kong}
  X.,  2014, \mn@doi [\mnras] {10.1093/mnras/stu1605}, \href
  {https://ui.adsabs.harvard.edu/abs/2014MNRAS.444.2525C} {444, 2525}

\bibitem[\protect\citeauthoryear{{Choi}, {Dotter}, {Conroy}, {Cantiello},
  {Paxton}  \& {Johnson}}{{Choi} et~al.}{2016}]{Choi2016}
{Choi} J.,  {Dotter} A.,  {Conroy} C.,  {Cantiello} M.,  {Paxton} B.,
  {Johnson} B.~D.,  2016, \mn@doi [\apj] {10.3847/0004-637X/823/2/102}, \href
  {https://ui.adsabs.harvard.edu/abs/2016ApJ...823..102C} {823, 102}

\bibitem[\protect\citeauthoryear{{Cutri} et~al.,}{{Cutri}
  et~al.}{2003}]{Cutri2003}
{Cutri} R.~M.,  et~al., 2003, VizieR Online Data Catalog, \href
  {https://ui.adsabs.harvard.edu/abs/2003yCat.2246....0C} {p. II/246}

\bibitem[\protect\citeauthoryear{{Da Rio} \& {Robberto}}{{Da Rio} \&
  {Robberto}}{2012}]{DaRio2012}
{Da Rio} N.,  {Robberto} M.,  2012, \mn@doi [\aj]
  {10.1088/0004-6256/144/6/176}, \href
  {https://ui.adsabs.harvard.edu/abs/2012AJ....144..176D} {144, 176}

\bibitem[\protect\citeauthoryear{{Degl'Innocenti}, {Prada Moroni}, {Marconi}
  \& {Ruoppo}}{{Degl'Innocenti} et~al.}{2008}]{DeglInnocenti2008}
{Degl'Innocenti} S.,  {Prada Moroni} P.~G.,  {Marconi} M.,   {Ruoppo} A.,
  2008, \mn@doi [\apss] {10.1007/s10509-007-9560-2}, \href
  {https://ui.adsabs.harvard.edu/abs/2008Ap&SS.316...25D} {316, 25}

\bibitem[\protect\citeauthoryear{{Dotter}}{{Dotter}}{2016}]{Dotter2016}
{Dotter} A.,  2016, \mn@doi [\apjs] {10.3847/0067-0049/222/1/8}, \href
  {https://ui.adsabs.harvard.edu/abs/2016ApJS..222....8D} {222, 8}

\bibitem[\protect\citeauthoryear{{Dotter}, {Chaboyer}, {Jevremovi{\'c}},
  {Kostov}, {Baron}  \& {Ferguson}}{{Dotter} et~al.}{2008}]{Dotter2008}
{Dotter} A.,  {Chaboyer} B.,  {Jevremovi{\'c}} D.,  {Kostov} V.,  {Baron} E.,
  {Ferguson} J.~W.,  2008, \mn@doi [\apjs] {10.1086/589654}, \href
  {https://ui.adsabs.harvard.edu/abs/2008ApJS..178...89D} {178, 89}

\bibitem[\protect\citeauthoryear{{Douglas}, {Ag{\"u}eros}, {Covey}, {Cargile},
  {Barclay}, {Cody}, {Howell}  \& {Kopytova}}{{Douglas}
  et~al.}{2016}]{Douglas2016}
{Douglas} S.~T.,  {Ag{\"u}eros} M.~A.,  {Covey} K.~R.,  {Cargile} P.~A.,
  {Barclay} T.,  {Cody} A.,  {Howell} S.~B.,   {Kopytova} T.,  2016, \mn@doi
  [\apj] {10.3847/0004-637X/822/1/47}, \href
  {https://ui.adsabs.harvard.edu/abs/2016ApJ...822...47D} {822, 47}

\bibitem[\protect\citeauthoryear{{Douglas}, {Curtis}, {Ag{\"u}eros}, {Cargile},
  {Brewer}, {Meibom}  \& {Jansen}}{{Douglas} et~al.}{2019}]{Douglas2019}
{Douglas} S.~T.,  {Curtis} J.~L.,  {Ag{\"u}eros} M.~A.,  {Cargile} P.~A.,
  {Brewer} J.~M.,  {Meibom} S.,   {Jansen} T.,  2019, \mn@doi [\apj]
  {10.3847/1538-4357/ab2468}, \href
  {https://ui.adsabs.harvard.edu/abs/2019ApJ...879..100D} {879, 100}

\bibitem[\protect\citeauthoryear{{Feiden}}{{Feiden}}{2016}]{Feiden2016}
{Feiden} G.~A.,  2016, \mn@doi [\aap] {10.1051/0004-6361/201527613}, \href
  {https://ui.adsabs.harvard.edu/abs/2016A&A...593A..99F} {593, A99}

\bibitem[\protect\citeauthoryear{{Feiden} \& {Chaboyer}}{{Feiden} \&
  {Chaboyer}}{2014}]{Feiden2014}
{Feiden} G.~A.,  {Chaboyer} B.,  2014, \mn@doi [\apj]
  {10.1088/0004-637X/789/1/53}, \href
  {https://ui.adsabs.harvard.edu/abs/2014ApJ...789...53F} {789, 53}

\bibitem[\protect\citeauthoryear{GAIA-Collaboration}{GAIA-Collaboration}{2020}]{gaia_edr3lite}
GAIA-Collaboration 2020, Gaia eDR3 source catalogue "light", {VO} resource
  provided by the {GAVO} Data Center, \url
  {https://dc.zah.uni-heidelberg.de/tableinfo/gaia.edr3lite}

\bibitem[\protect\citeauthoryear{{Gaia Collaboration} et~al.,}{{Gaia
  Collaboration} et~al.}{2016}]{GAIA2016}
{Gaia Collaboration} et~al., 2016, \mn@doi [\aap]
  {10.1051/0004-6361/201629272}, \href
  {https://ui.adsabs.harvard.edu/abs/2016A&A...595A...1G} {595, A1}

\bibitem[\protect\citeauthoryear{{Gaia Collaboration} et~al.,}{{Gaia
  Collaboration} et~al.}{2018}]{GAIA2018}
{Gaia Collaboration} et~al., 2018, \mn@doi [\aap]
  {10.1051/0004-6361/201833051}, \href
  {https://ui.adsabs.harvard.edu/abs/2018A&A...616A...1G} {616, A1}

\bibitem[\protect\citeauthoryear{{Gaia Collaboration} et~al.,}{{Gaia
  Collaboration} et~al.}{2021}]{GAIA_Smart2021A}
{Gaia Collaboration} et~al., 2021, \mn@doi [\aap]
  {10.1051/0004-6361/202039498}, \href
  {https://ui.adsabs.harvard.edu/abs/2021A&A...649A...6G} {649, A6}

\bibitem[\protect\citeauthoryear{{Girardi}, {Bertelli}, {Bressan}, {Chiosi},
  {Groenewegen}, {Marigo}, {Salasnich}  \& {Weiss}}{{Girardi}
  et~al.}{2002}]{Giradi2002}
{Girardi} L.,  {Bertelli} G.,  {Bressan} A.,  {Chiosi} C.,  {Groenewegen}
  M.~A.~T.,  {Marigo} P.,  {Salasnich} B.,   {Weiss} A.,  2002, \mn@doi [\aap]
  {10.1051/0004-6361:20020612}, \href
  {https://ui.adsabs.harvard.edu/abs/2002A&A...391..195G} {391, 195}

\bibitem[\protect\citeauthoryear{{Goldman} et~al.,}{{Goldman}
  et~al.}{2013}]{Goldmann2013}
{Goldman} B.,  et~al., 2013, \mn@doi [\aap] {10.1051/0004-6361/201321727},
  \href {https://ui.adsabs.harvard.edu/abs/2013A&A...559A..43G} {559, A43}

\bibitem[\protect\citeauthoryear{{Gossage}, {Conroy}, {Dotter}, {Choi},
  {Rosenfield}, {Cargile}  \& {Dolphin}}{{Gossage} et~al.}{2018}]{Gossage2018}
{Gossage} S.,  {Conroy} C.,  {Dotter} A.,  {Choi} J.,  {Rosenfield} P.,
  {Cargile} P.,   {Dolphin} A.,  2018, \mn@doi [\apj]
  {10.3847/1538-4357/aad0a0}, \href
  {https://ui.adsabs.harvard.edu/abs/2018ApJ...863...67G} {863, 67}

\bibitem[\protect\citeauthoryear{{H{\o}g} et~al.,}{{H{\o}g}
  et~al.}{2000}]{Hog2000}
{H{\o}g} E.,  et~al., 2000, \aap, \href
  {https://ui.adsabs.harvard.edu/abs/2000A&A...355L..27H} {355, L27}

\bibitem[\protect\citeauthoryear{{Hormuth}, {Hippler}, {Brandner}, {Wagner}  \&
  {Henning}}{{Hormuth} et~al.}{2008}]{Hormuth2008}
{Hormuth} F.,  {Hippler} S.,  {Brandner} W.,  {Wagner} K.,   {Henning} T.,
  2008, in {McLean} I.~S.,  {Casali} M.~M.,  eds,  Society of Photo-Optical
  Instrumentation Engineers (SPIE) Conference Series Vol. 7014, Ground-based
  and Airborne Instrumentation for Astronomy II. p. 701448 (\mn@eprint {arXiv}
  {0807.0497}), \mn@doi{10.1117/12.787384}

\bibitem[\protect\citeauthoryear{{Ireland} \& {Browning}}{{Ireland} \&
  {Browning}}{2018}]{Ireland2018}
{Ireland} L.~G.,  {Browning} M.~K.,  2018, \mn@doi [\apj]
  {10.3847/1538-4357/aab3da}, \href
  {https://ui.adsabs.harvard.edu/abs/2018ApJ...856..132I} {856, 132}

\bibitem[\protect\citeauthoryear{{Kopytova}, {Brandner}, {Tognelli}, {Prada
  Moroni}, {Da Rio}, {R{\"o}ser}  \& {Schilbach}}{{Kopytova}
  et~al.}{2016}]{Kopytova2016}
{Kopytova} T.~G.,  {Brandner} W.,  {Tognelli} E.,  {Prada Moroni} P.~G.,  {Da
  Rio} N.,  {R{\"o}ser} S.,   {Schilbach} E.,  2016, \mn@doi [\aap]
  {10.1051/0004-6361/201527044}, \href
  {http://esoads.eso.org/abs/2016A%26A...585A...7K} {585, A7}

\bibitem[\protect\citeauthoryear{{Krumholz}, {McKee}  \&
  {Bland-Hawthorn}}{{Krumholz} et~al.}{2019}]{Krumholz2019}
{Krumholz} M.~R.,  {McKee} C.~F.,   {Bland-Hawthorn} J.,  2019, \mn@doi [\araa]
  {10.1146/annurev-astro-091918-104430}, \href
  {https://ui.adsabs.harvard.edu/abs/2019ARA&A..57..227K} {57, 227}

\bibitem[\protect\citeauthoryear{{Lindegren} et~al.,}{{Lindegren}
  et~al.}{2021}]{Lindegren2021}
{Lindegren} L.,  et~al., 2021, \mn@doi [\aap] {10.1051/0004-6361/202039709},
  \href {https://ui.adsabs.harvard.edu/abs/2021A&A...649A...2L} {649, A2}

\bibitem[\protect\citeauthoryear{{MacDonald} \& {Mullan}}{{MacDonald} \&
  {Mullan}}{2014}]{MacDonald2014}
{MacDonald} J.,  {Mullan} D.~J.,  2014, \mn@doi [\apj]
  {10.1088/0004-637X/787/1/70}, \href
  {https://ui.adsabs.harvard.edu/abs/2014ApJ...787...70M} {787, 70}

\bibitem[\protect\citeauthoryear{{Marigo}, {Girardi}, {Bressan}, {Groenewegen},
  {Silva}  \& {Granato}}{{Marigo} et~al.}{2008}]{Marigo2008}
{Marigo} P.,  {Girardi} L.,  {Bressan} A.,  {Groenewegen} M.~A.~T.,  {Silva}
  L.,   {Granato} G.~L.,  2008, \mn@doi [\aap] {10.1051/0004-6361:20078467},
  \href {https://ui.adsabs.harvard.edu/abs/2008A&A...482..883M} {482, 883}

\bibitem[\protect\citeauthoryear{{Mermilliod}}{{Mermilliod}}{1995}]{Mermilliod1995}
{Mermilliod} J.-C.,  1995, in {Egret} D.,  {Albrecht} M.~A.,  eds, ~ASSL Vol.
  203, Information \& On-Line Data in Astronomy. p.~127,
  \mn@doi{10.1007/978-94-011-0397-8\_12}

\bibitem[\protect\citeauthoryear{{Paxton}, {Bildsten}, {Dotter}, {Herwig},
  {Lesaffre}  \& {Timmes}}{{Paxton} et~al.}{2011}]{Paxton2011}
{Paxton} B.,  {Bildsten} L.,  {Dotter} A.,  {Herwig} F.,  {Lesaffre} P.,
  {Timmes} F.,  2011, \mn@doi [\apjs] {10.1088/0067-0049/192/1/3}, \href
  {https://ui.adsabs.harvard.edu/abs/2011ApJS..192....3P} {192, 3}

\bibitem[\protect\citeauthoryear{{Paxton} et~al.,}{{Paxton}
  et~al.}{2013}]{Paxton2013}
{Paxton} B.,  et~al., 2013, \mn@doi [\apjs] {10.1088/0067-0049/208/1/4}, \href
  {https://ui.adsabs.harvard.edu/abs/2013ApJS..208....4P} {208, 4}

\bibitem[\protect\citeauthoryear{{Paxton} et~al.,}{{Paxton}
  et~al.}{2015}]{Paxton2015}
{Paxton} B.,  et~al., 2015, \mn@doi [\apjs] {10.1088/0067-0049/220/1/15}, \href
  {https://ui.adsabs.harvard.edu/abs/2015ApJS..220...15P} {220, 15}

\bibitem[\protect\citeauthoryear{{Paxton} et~al.,}{{Paxton}
  et~al.}{2018}]{Paxton2018}
{Paxton} B.,  et~al., 2018, \mn@doi [\apjs] {10.3847/1538-4365/aaa5a8}, \href
  {https://ui.adsabs.harvard.edu/abs/2018ApJS..234...34P} {234, 34}

\bibitem[\protect\citeauthoryear{{Perryman} et~al.,}{{Perryman}
  et~al.}{1998}]{Perryman1998}
{Perryman} M.~A.~C.,  et~al., 1998, \aap, \href
  {https://ui.adsabs.harvard.edu/abs/1998A&A...331...81P} {331, 81}

\bibitem[\protect\citeauthoryear{{Riello} et~al.,}{{Riello}
  et~al.}{2021}]{Riello2021}
{Riello} M.,  et~al., 2021, \mn@doi [\aap] {10.1051/0004-6361/202039587}, \href
  {https://ui.adsabs.harvard.edu/abs/2021A&A...649A...3R} {649, A3}

\bibitem[\protect\citeauthoryear{{R{\"o}ser}, {Demleitner}  \&
  {Schilbach}}{{R{\"o}ser} et~al.}{2010}]{Roeser2010}
{R{\"o}ser} S.,  {Demleitner} M.,   {Schilbach} E.,  2010, \mn@doi [\aj]
  {10.1088/0004-6256/139/6/2440}, \href
  {https://ui.adsabs.harvard.edu/abs/2010AJ....139.2440R} {139, 2440}

\bibitem[\protect\citeauthoryear{{R{\"o}ser}, {Schilbach}, {Piskunov},
  {Kharchenko}  \& {Scholz}}{{R{\"o}ser} et~al.}{2011}]{Roeser11}
{R{\"o}ser} S.,  {Schilbach} E.,  {Piskunov} A.~E.,  {Kharchenko} N.~V.,
  {Scholz} R.-D.,  2011, \mn@doi [A\&A] {10.1051/0004-6361/201116948}, \href
  {http://adsabs.harvard.edu/abs/2011A%26A...531A..92R} {531, A92}

\bibitem[\protect\citeauthoryear{{Somers} \& {Pinsonneault}}{{Somers} \&
  {Pinsonneault}}{2015}]{Somers2015a}
{Somers} G.,  {Pinsonneault} M.~H.,  2015, \mn@doi [\mnras]
  {10.1093/mnras/stv630}, \href
  {https://ui.adsabs.harvard.edu/abs/2015MNRAS.449.4131S} {449, 4131}

\bibitem[\protect\citeauthoryear{{Somers} \& {Stassun}}{{Somers} \&
  {Stassun}}{2017}]{Somers2017}
{Somers} G.,  {Stassun} K.~G.,  2017, \mn@doi [\aj]
  {10.3847/1538-3881/153/3/101}, \href
  {https://ui.adsabs.harvard.edu/abs/2017AJ....153..101S} {153, 101}

\bibitem[\protect\citeauthoryear{{Stevenson}}{{Stevenson}}{1979}]{Stevenson1979}
{Stevenson} D.~J.,  1979, \mn@doi [Geophysical and Astrophysical Fluid
  Dynamics] {10.1080/03091927908242681}, \href
  {https://ui.adsabs.harvard.edu/abs/1979GApFD..12..139S} {12, 139}

\bibitem[\protect\citeauthoryear{{Tognelli}, {Prada Moroni}  \&
  {Degl'Innocenti}}{{Tognelli} et~al.}{2011}]{Tognelli2011}
{Tognelli} E.,  {Prada Moroni} P.~G.,   {Degl'Innocenti} S.,  2011, \mn@doi
  [\aap] {10.1051/0004-6361/200913913}, \href
  {https://ui.adsabs.harvard.edu/abs/2011A&A...533A.109T} {533, A109}

\bibitem[\protect\citeauthoryear{{Tognelli}, {Degl'Innocenti}  \& {Prada
  Moroni}}{{Tognelli} et~al.}{2012}]{Tognelli2012}
{Tognelli} E.,  {Degl'Innocenti} S.,   {Prada Moroni} P.~G.,  2012, \mn@doi
  [\aap] {10.1051/0004-6361/201219111}, \href
  {https://ui.adsabs.harvard.edu/abs/2012A&A...548A..41T} {548, A41}

\bibitem[\protect\citeauthoryear{{Tognelli}, {Dell'Omodarme}, {Valle}, {Prada
  Moroni}  \& {Degl'Innocenti}}{{Tognelli} et~al.}{2021}]{Tognelli2021}
{Tognelli} E.,  {Dell'Omodarme} M.,  {Valle} G.,  {Prada Moroni} P.~G.,
  {Degl'Innocenti} S.,  2021, \mn@doi [\mnras] {10.1093/mnras/staa3686}, \href
  {https://ui.adsabs.harvard.edu/abs/2021MNRAS.501..383T} {501, 383}

\bibitem[\protect\citeauthoryear{{de Bruijne}, {Hoogerwerf}  \& {de Zeeuw}}{{de
  Bruijne} et~al.}{2001}]{deBruijne2001}
{de Bruijne} J.~H.~J.,  {Hoogerwerf} R.,   {de Zeeuw} P.~T.,  2001, \mn@doi
  [\aap] {10.1051/0004-6361:20000410}, \href
  {https://ui.adsabs.harvard.edu/abs/2001A&A...367..111D} {367, 111}

\bibitem[\protect\citeauthoryear{{van Saders} \& {Pinsonneault}}{{van Saders}
  \& {Pinsonneault}}{2012}]{vanSaders2012}
{van Saders} J.~L.,  {Pinsonneault} M.~H.,  2012, \mn@doi [\apj]
  {10.1088/0004-637X/751/2/98}, \href
  {https://ui.adsabs.harvard.edu/abs/2012ApJ...751...98V} {751, 98}

\makeatother
\end{thebibliography}


\bsp	
\label{lastpage}
\end{document}